\documentclass[aip,reprint]{revtex4-1}
\usepackage{amssymb}
\usepackage{amsfonts}
\usepackage{amsmath}
\usepackage{color}
\usepackage{graphicx}

\begin{document}




\title{Tracking Emission Rate Dynamics of NV Centers in Nanodiamonds}

\author{F A Inam}
\affiliation{
MQ Photonics Research Centre, Department of Physics and Astronomy, Macquarie University, North Ryde, New South Wales 2109, Australia}
\affiliation{
Centre for Quantum Science and Technology, Department of Physics and Astronomy, Macquarie University, North Ryde, New South Wales 2109, Australia}
\affiliation{
ARC Centre of Excellence for Engineered Quantum Systems (EQuS), Department of Physics and Astronomy, Macquarie University, North Ryde, New South Wales 2109, Australia}

\author{A M Edmonds}
\affiliation{
MQ Photonics Research Centre, Department of Physics and Astronomy, Macquarie University, North Ryde, New South Wales 2109, Australia}
\affiliation{
Centre for Quantum Science and Technology, Department of Physics and Astronomy, Macquarie University, North Ryde, New South Wales 2109, Australia}
\affiliation{
ARC Centre of Excellence for Engineered Quantum Systems (EQuS), Department of Physics and Astronomy, Macquarie University, North Ryde, New South Wales 2109, Australia}

\author{M J Steel}
\email{michael.steel@mq.edu.au}
\affiliation{
MQ Photonics Research Centre, Department of Physics and Astronomy, Macquarie University, North Ryde, New South Wales 2109, Australia}
\affiliation{
Centre for Quantum Science and Technology, Department of Physics and Astronomy, Macquarie University, North Ryde, New South Wales 2109, Australia}

\author{S Castelletto}
\affiliation{
MQ Photonics Research Centre, Department of Physics and Astronomy, Macquarie University, North Ryde, New South Wales 2109, Australia}
\affiliation{
Centre for Quantum Science and Technology, Department of Physics and Astronomy, Macquarie University, North Ryde, New South Wales 2109, Australia}
\affiliation{
ARC Centre of Excellence for Engineered Quantum Systems (EQuS), Department of Physics and Astronomy, Macquarie University, North Ryde, New South Wales 2109, Australia}

\begin{abstract}
Spontaneous emission from crystal centers is influenced by both the photonic local density of states and non-radiative processes. Here we monitor the spontaneous emission of single nitrogen vacancy (NV) centers as their host diamond is reduced in size from a large monolithic crystal to a nanocrystal by successive cycles of oxidation. The size reduction induces a quenching of the NV radiative emission. New non-radiative channels lead to a decrease of the fluorescence intensity and the excited state lifetime. In one case we observe the onset of blinking which may provide a route to understand these additional non-radiative decay channels.
\end{abstract}


\maketitle 

Negatively charged nitrogen-vacancy (NV) centers in diamond have attracted wide interest in interdisciplinary research fields, ranging from quantum technologies \cite{Childress13_MRS} and high-resolution magnetometry \cite{Wrachtrup13_MRS} to biomedical imaging and tracking \cite{LeSage13}. Their unique optical and spin properties have revealed some important functionalities such as bright, stable room-temperature single photon emission \cite{Loncar13_MRS}, high resolution room temperature magnetometry \cite{Maze08, Wrachtrup13_MRS}, a suitability for sub-diffraction limited  microscopy \cite{Maletinsky12, Rittweger09}, and high biocompatibility for labelling \cite{LeSage13}. Many of the sensing applications  are highly dependent on the separation between the NV dipole and the surrounding electric and/or magnetic fields \cite{Wrachtrup13_MRS}. Therefore these  applications require the NV to be either enclosed inside very small nanodiamonds or created very close to the bulk diamond surface.


However, it is known that NV center spontaneous emission properties deteriorate dramatically in terms of spectral diffusion and photo-stability on moving from a monolithic ultra-pure single crystal diamond host to its counterpart nanosized crystal.\cite{Tamarat06} Compared to the bulk, both high pressure high temperature (HPHT) and chemical vapour deposition (CVD) nanodiamonds show a larger spectral diffusion of NV spontaneous emission due to the presence of other impurities associated with growth methods or contamination in milling procedures.\cite{Santori10}
Recent progress in identifying and characterizing the deterioration of NV spontaneous emission from bulk diamond to nanodiamonds \cite{Tamarat06, Santori10, Waldherr11, Siyushev12} shows it cannot be explained by purely electromagnetic  arguments.\cite{Inam11,Mohtashami13} Several recent studies by ourselves and others point to the existence of non-radiative decay mechanisms related to the NV emission inside the nanodiamond crystals.\cite{Inam11,Mohtashami13,Inam13}
Following our work revealing an incompatibility between
observed NV centre decay rates from nanodiamonds ($\sim$50 nm)
and calculations assuming unit quantum efficiency (QE)\cite{Inam11}, a recent study showed evidence for a large distribution in the NV QE related to the nanodiamond size.\cite{Mohtashami13} In diamond crystals of 
order 100~nm size, this work found a broad distribution of NV QE between 10\% to 90\%, while for smaller crystals ($\sim$25~nm), QE was restricted to just 0--20\%. Very recently, by studying NV emission in a near substrate-free aerogel environment of refractive-index close to unity (\textit{n}$\sim$1.05), we estimated an upper-bound for the mean QE in nanodiamond ($\sim$50~nm) as $\bar{\eta}$ = 0.7 \cite{Inam13}. The commonly-used assumption in earlier studies \cite{Gruber97, Schietinger09nl} 
of near-unity quantum efficiency is thus no longer appropriate.

Studies on ``single-digit'' (diameter below 10~nm) nanodiamonds have also revealed the strong influence of the local environment in drastically reducing the NV spontaneous emission. The outer graphitic shell enclosing the diamond crystal has been found to cause quenching of the NV emission \cite{Smith10}. In a specific type of nanodiamonds, obtained via detonation in the presence of carbon, NV has been found to possess unstable fluorescence emission inducing photoluminescence intermittency (blinking) \cite{Bradac2010}. More recently, NV photoluminescence intermittency of single defects has also been observed in HPHT nanodiamonds \cite{Bradac13}. 

Since the potential applications of nanodiamonds in biology and magnetometry strongly rely on NV dipole emission from very small diamonds, and since biological applications also require their surface functionalisation, it is critical to identify fully  the conditions and mechanisms which suppress NV spontaneous emission or induce blinking behavior. 

In this work we aim to probe the influences on the NV decay rate and emission arising from the NV center proximity to the diamond's surface. To this end, we dynamically studied the spontaneous emission from the same single NV centers while shrinking the size of the host diamond crystals via air oxidation from large diamond crystals down to nanodiamond scale. \cite{Gaebel12} By spreading the nanodiamonds ($\sim$500 nm) on a laser inscribed patterned glass coverslip, we could identify individual diamonds  and track the spontaneous emission from single nanodiamond NV centers across successive cycles of  oxidation. At each stage,  we measured the spontaneous emission decay rate, saturation count rate and photon statistics of the single NV centers. This continuous tracking of the NV center spontaneous emission against the host diamond crystal size provides a new level of interrogation and insight into the NV spontaneous emission. 

Our technique serves to distinguish the electromagnetic influence of the local diamond crystal on the emission properties of the NV from the other non-electromagnetic contributions affecting the emission, possibly induced by the size reduction procedure.
As the size of the host crystal shrinks to the nano-regime, the NV emission rate is predicted to be highly suppressed by the Lorenz-Lorentz reduction of the local electric field inside the sub-wavelength dielectric crystal.\cite{Chew87,Inam11,Schniepp02} 
On electromagnetic grounds then, the high refractive index of diamond should favor a strong electromagnetic suppression of the count rate and an increase in the decay lifetime for smaller crystals. Any departure from this behavior is evidence for changes in the non-radiative dynamics.

To prepare our samples, we used a commercially available polishing grade HPHT diamond (type Ib) solution in milliQ water (VanMoppes, SYP-GAF 0.5-1~$\mu$m, 60~mg/ml) that was first ultra-sonicated with  a horn-type  sonotrode for 1~hr to remove aggregation. 
  A dispersed solution of diamonds in water was sprayed onto a glass cover slip which was laser inscribed with a 5 $\times$ 5 grid of squares of side length 50~$\mu$m. This grid facilitated identification of the same diamond crystals over consecutive oxidation steps.

\begin{figure}[t!]  
\centering
\includegraphics[width=8.5cm]{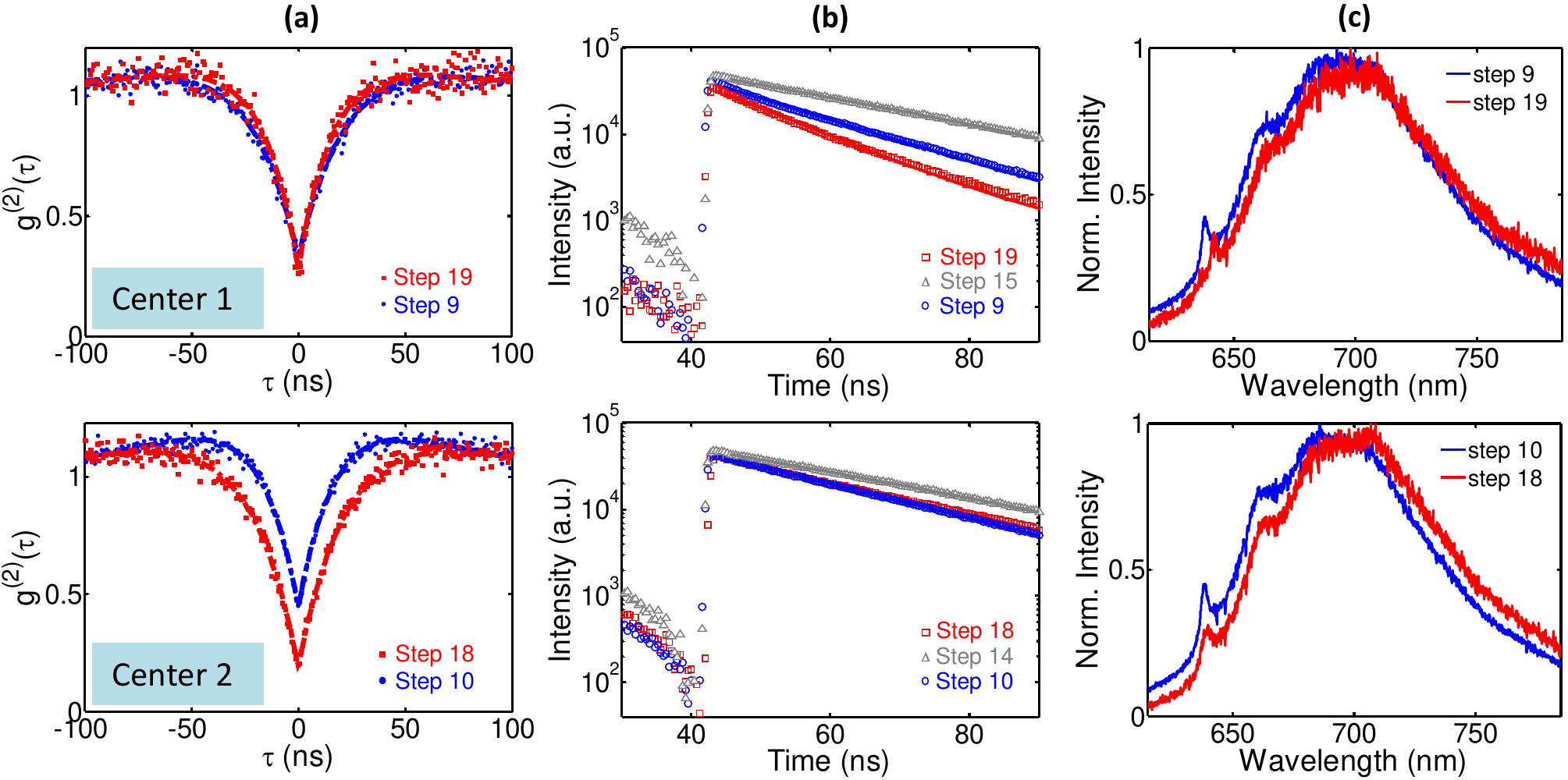}
\caption[]{(a) Second order auto-correlation, b) time-resolved fluorescence decay and (c) fluorescence spectrum curves for the two single NV centers (1 and 2) corresponding to the initial (blue) and final (red) oxidation steps of the host crystal. For fluorescence decay, an intermediate (grey) oxidation step curve is also shown. The $g^{(2)}$ curves are fitted to the 3-level model \cite{Kurtsiefer00}, and the fluorescence lifetime curves are fitted to a single exponential decay \cite{Tisler09}. }
\label{Initial_final_SE_sizered}
\end{figure}

The sample fluorescence was simultaneously measured with a confocal scanning fluorescence microscope (100$\times$ oil immersion objective lens, NA 1.4), excited with a 532 nm CW diode pumped solid-state laser, and a commercial atomic force microscope (AFM)\cite{Bradac09}. A spectrometer with a cooled CCD was used to characterize the luminescence and a Hanbury-Brown and Twiss (HBT) interferometer with single-photon-sensitive avalanche photodiodes was used to measure the photon statistics \cite{Kurtsiefer00}. Photon counting and correlation was carried out using a time-correlated single-photon-counting (TCSPC) module. The excited state NV$^-$ lifetime  was determined using time-resolved fluorescence measurements with pulsed 532 nm laser excitation \cite{Tisler09} and collection of the NV emission only above 650 nm.\cite{Beha12} For this purpose a long-pass 650 nm filter was placed before the photodetectors (but not the spectrometer). 
As a result, though the NV$^-$ ZPL~(637 nm) appears in the spectrum (Fig.~\ref{Initial_final_SE_sizered}), there is no influence from the NV$^0$ emission, which is restricted to wavelengths below 630 nm\cite{Beha12}, on the measured lifetimes of the NV$^-$ centers. 

\begin{figure}[t!]  
\includegraphics[width=8cm]{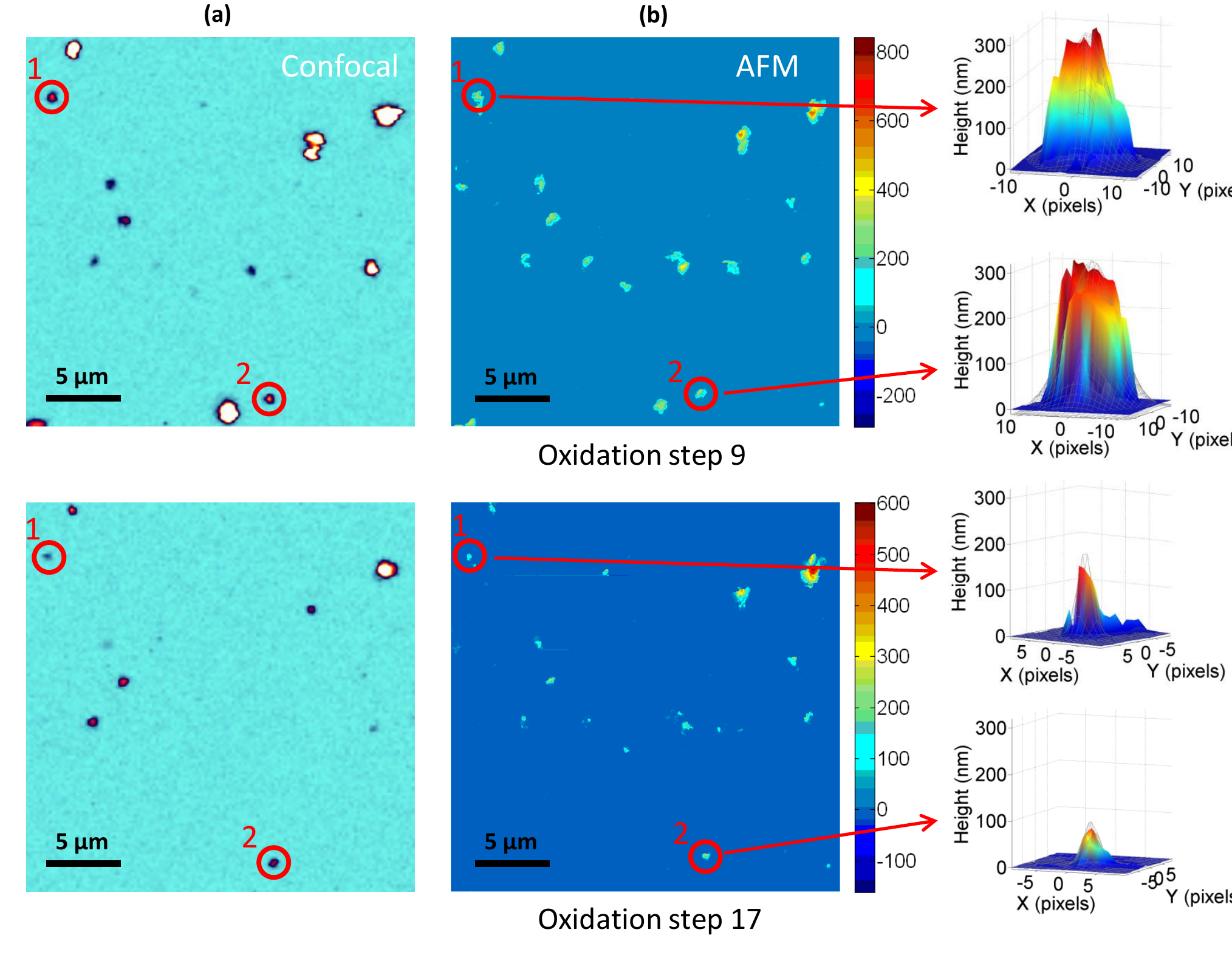}
\caption[]{(a) Confocal and (b) AFM images of the sample area corresponding to the two single NV centers for two oxidation steps (9 and 17). The 3D AFM images represents the height of the two nanodiamond hosts.}
\label{Afm_confocal_sizered}
\end{figure}

The oxidation process for the diamond sample 
was carried out in a furnace in air at atmospheric pressure. For the initial step, the sample was oxidized for a duration of 1 hr at 500$^\circ$C to remove the surface graphitic phase \cite{Smith10}. The subsequent oxidation cycles were performed at 580$^\circ$C for a duration of 4-5 hrs each.
AFM studies determined the initial average height of these diamond crystals to be around 500~nm. After each oxidation cycle, simultaneous AFM and fluorescence studies were performed to isolate the single NV centers. Only single centers were studied to remove averaging effects due to multiple centers. Single centers were identified from the measurement of the second order intensity correlation function $g^{(2)}(\tau)$, selecting only those diamonds displaying $g^{(2)}(0)<$ 0.5. This proved challenging as the probability of finding single NV centers inside these large diamond crystals is estimated to be well below 1$\%$ \cite{Bradac09}. Therefore the sample area of one 50~$\mu$m grid-square which was selected for this study was the one initially having the maximum number of centers (22) displaying some anti-bunching behavior ($g^{(2)}(0)<$ 1) in the intensity correlation curve. For the whole study we carried out a total of 20 oxidation cycles and measured more than 200 intensity correlation curve for centers in the sample area. In the complete process we found only four single centers. For two centers (Fig. \ref{Afm_confocal_sizered}), we were able to track the spontaneous emission properties for a significant range of diamond crystal height (for 9 and 11 oxidation cycles respectively). 
Diamond 1 reached $g^{(2)}(0)<$ 0.5 in the 9th oxidation step and lasted till the 19th step; diamond 2 showed single center behavior between oxidation steps 10-18. The remaining two centers evaporated only a few oxidation steps (3 and 4 respectively) after they began to show single center characteristics. 
Figure~\ref{Afm_confocal_sizered} shows the confocal scan images of the two long lived centers (marked with red circles) along with their corresponding AFM height images at early (upper row) and late (lower) steps in the size reduction procedure.

\begin{figure}[t!]  
\includegraphics[width=8.5cm]{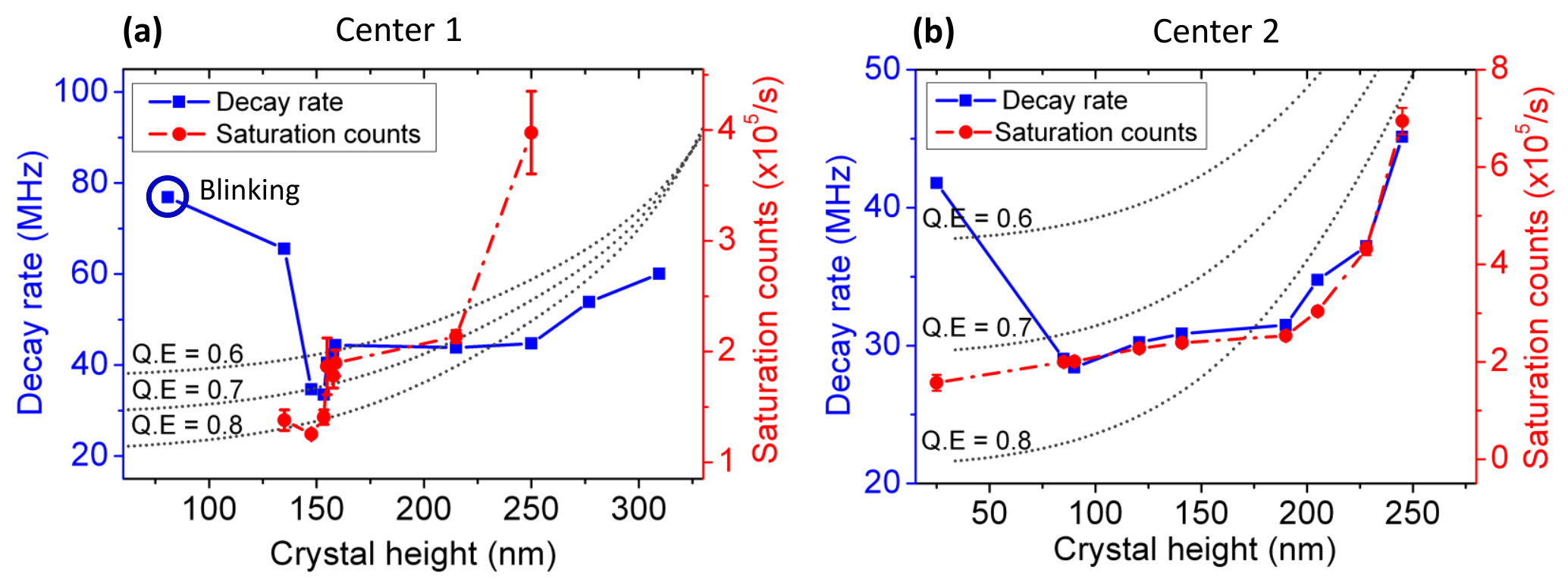}
\caption[]{Total decay rate (solid blue) and the photon saturation counts (dashed red) as a function of the measured crystal heights for the two single NV centers (a) center~1 (b) center~2. For center~1, the last oxidation step emission rate represents the emission rate for the blinking state. Smooth dotted curves shows the analytic decay rates for the diamond sphere NV centers emitting with different quantum efficiencies.
}
\label{Decay_sizered}
\end{figure}

Figure~\ref{Initial_final_SE_sizered} shows the optical properties (intensity correlation function, time-resolved fluorescence decay curve, and emission spectrum) for the two NV centers at the first and last oxidation steps for which they showed single center behavior. 
We note that anti-bunching decay curves are often used to determine the emission lifetime,\cite{Kurtsiefer00} and
in this case it may appear that the initial and final anti-bunching curves for the two centers are inconsistent with
the lifetime changes observed in Fig.~\ref{Initial_final_SE_sizered}b.
In fact, this comparison is only  valid when the excitation power for the comparative cases is the same.\cite{Kurtsiefer00} In our measurements, as the photon counts for the subsequent steps were continuously decreasing (Fig.~\ref{Decay_sizered}), the photon-autocorrelation curves for the steps shown in Fig.~\ref{Initial_final_SE_sizered}a were for different excitation powers.


Figure~\ref{Decay_sizered} shows the decay rate (inverse decay lifetime) as a function
of the crystal height (blue). (The temporal evolution is towards the left.)
For these centers, the decay rate initially decreased as the size of the host diamond crystal was reduced. We associate this with the expected suppression of the local density of states available for the emission of a photon due to the Lorenz-Lorentz reduction of the local electric field inside the high refractive index diamond crystal\cite{Chew87}.
For comparison, we show the predicted decay rates (dotted gray lines) for an NV at the centre of an ideal spherical diamond for several values of quantum efficiency.\cite{Chew87}  As is well known, these rates decrease with decreasing size
to a constant value at small diameter $d \ll \lambda$, where the internal electric field is independent of the nanoparticle size.
Of course our system is not truly a sphere in vacuum but an irregularly shaped diamond crystal held on a coverslip.
We have also not included polarization-dependent effects due to the air-coverglass which are significant only for dipole-interface separation less than around 100 nm.\cite{Inam11}
Nevertheless, these curves provide a qualitative indication of the expected radiation dynamics
due to the changing electromagnetic local density of states inside the nanocrystals.
The similarity of the decay curves to the analytic QE=0.8 curves for diamonds above 200~nm suggests the LDOS effect
is dominant in this regime.
The suppression in the radiative emission is also evident from the corresponding reduction in the photon saturation counts (dashed red curves).

At intermediate crystal sizes ($\approx200$~nm), the decay
rate flattens out in a striking fashion, crossing the analytic curves to lower values of quantum efficiency and suggesting that
the non-radiative decay is becoming more significant.
As the diamond further shrinks to the nanoscale ($\lesssim150$~nm for center~1, and $\lesssim100$~nm for center~2), the pattern of the emission rate variation tends to reverse. The decay rate now increases sharply but the corresponding photon saturation counts do not show any increase. This change therefore represents the rapid appearance of 
new non-radiative decay mechanisms. This change in behavior is again highlighted by comparison with the analytic curves for varying quantum efficiencies.

Note that the reduction in the crystal height for center~1 (Fig. \ref{Decay_sizered}a) does not appear to follow a regular pattern (the blue squares denoting are not equally spaced along the crystal height axis). The polishing grade HPHT diamond crystals used in the experiments were of large crystal dimensions (0.5--1.0 $\mu$m) having a wide range of crystal geometries.  The reason for this non-uniformity in the size reduction may therefore be the anisotropy in the etch rate for different crystallographic planes as previously observed in CVD diamond samples \cite{Wolfer09} and natural diamond.\cite{Chu95} As the diamond shrinks, different crystal surfaces are exposed to the air for oxidation due to the irregularities in the crystal morphology.

\begin{figure}[t!]  
\centering
\includegraphics[width=8.5cm]{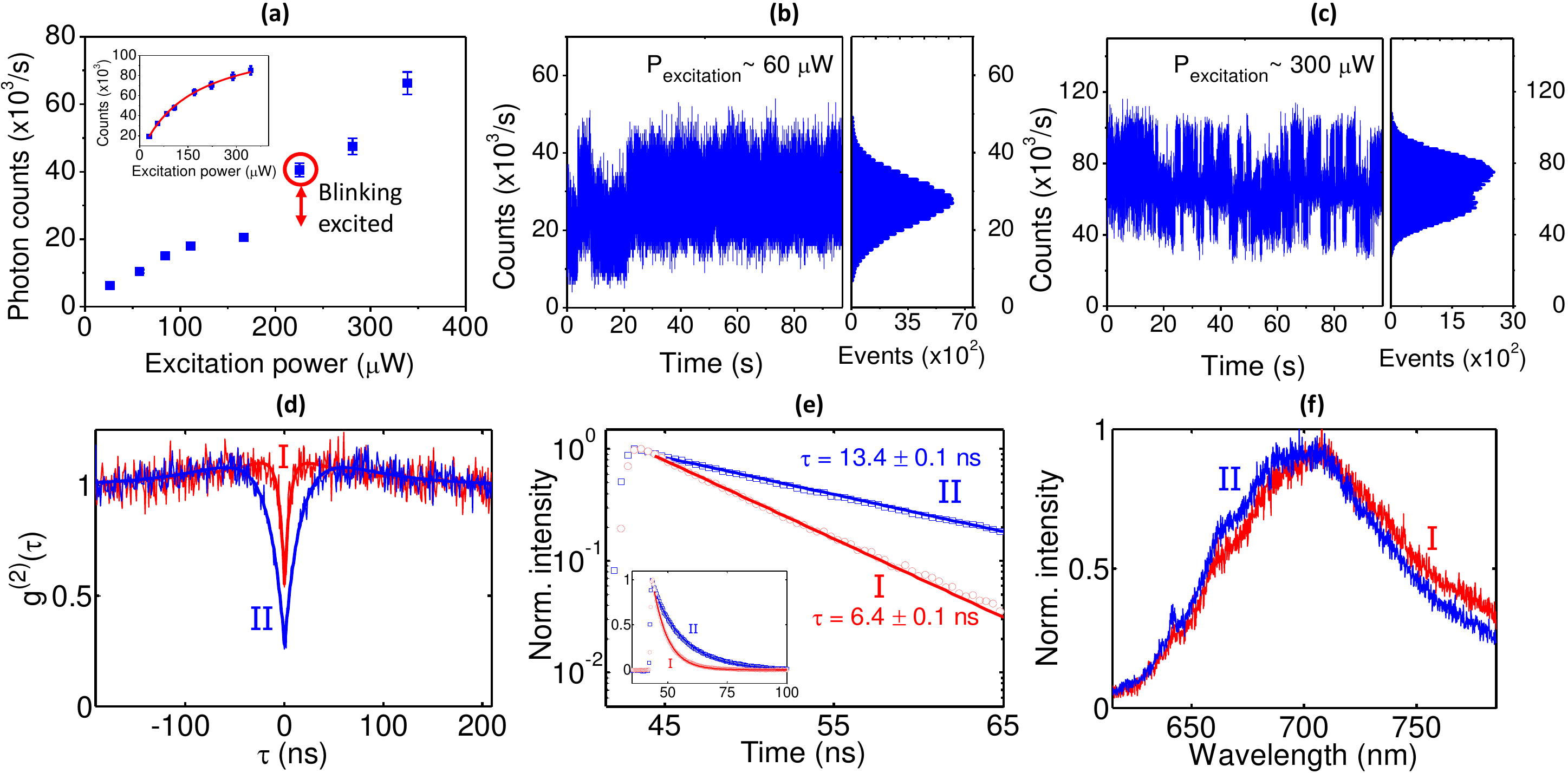}
\caption[]{
Blinking behavior of center~1. (a) Photon counts as a function of laser excitation power for the oxidation step when blinking was excited. (Inset) Photon saturation curve of the same center in the previous oxidation step (the red curve is a fit to the saturation model, $C = C_{\text{sat}} P/(P+P_{\text{sat}})$, where $P$ is the excitation power). Fluorescence time-trace for two different excitation powers (b) and (c) with the corresponding count-rate histograms. (d) Second order auto-correlation curves and (e) associated emission lifetimes corresponding to I (red) the low count state and II (blue), the predominantly high count blinking state. (f) Fluorescence spectrum of the center in the I (red) low-count state before the excitation of blinking, and the II state (blue) after the advent of blinking in a predominantly high count state at a low excitation power.}
\label{Blinking_sizered}
\end{figure}

In its last oxidation step, center~1 also showed evidence of intermittency in the luminescence (blinking behavior). Figure~\ref{Blinking_sizered} shows the saturation count rate (Fig. \ref{Blinking_sizered}a) and PL time traces (Fig. \ref{Blinking_sizered}(b,c)) of this center at two different optical excitation powers once blinking had begun. 
The blinking behavior was observed to be strongly dependent on the excitation laser power. In this oxidation step, the center was observed to be initially stable and the blinking behavior was only initiated on increasing the excitation power while measuring the center's photon saturation curve (Fig. \ref{Blinking_sizered}a). The center remained in its stable low count initial state up to an excitation power of 200 $\mu$W while displaying a typical saturation behavior (inset of Fig. \ref{Blinking_sizered}a (top) shows the saturation behavior of the same center in the previous oxidation step). On further increasing the laser power, the fluorescence count suddenly jumped to a higher rate and started blinking. Even on reducing the laser excitation power this behavior continued but the rate of blinking was observed to vary with the laser power, increasing with an increase in the laser excitation power (Fig. \ref{Blinking_sizered}(b,c)). It can also be seen from Fig.~\ref{Blinking_sizered}(b) that for low excitation power the center stays in the high count blinking state for the majority of time. 


Figure~\ref{Blinking_sizered}(d,e,f) shows the second order auto-correlation function (laser excitation power $\sim$100 $\mu$W), the time-resolved normalized fluorescence decay (pulsed laser excitation, power $\sim$8 $\mu$W) and the  fluorescence spectrum (laser excitation power $\sim$100 $\mu$W) curves for center~1 in the low count rate state before the blinking started (I), and after blinking commenced (II). The time-resolved fluorescence decay clearly shows a shorter emission lifetime for the center in its initial low-count state before blinking (I), compared to its emission lifetime in the blinking state (II) (where the center is expected to remain in the high count state for the greater fraction of time, considering the low excitation power). 

We also observed a red shift in the NV fluorescence spectrum for the emitter in its lower count state before blinking started (Fig. \ref{Blinking_sizered}f). (In this state, the decay process is expected to have a higher non-radiative decay contribution. This fact is also observed from the measured time-resolved fluorescence decay curves (Fig. \ref{Blinking_sizered}e), showing a comparatively much lower lifetime for the center in this state). This red-shift was also observed in both centers between the initial and the final oxidation steps as the non-radiative channels become evident (Fig. \ref{Initial_final_SE_sizered}c). This may suggest a correlation between the red shift and the advent of non-radiative decay channels for NV emission inside the nanocrystals. 

The blinking phenomenon in nanodiamond NV centers is still largely unclear. It is often attributed to fluctuating charges inside or at the surfaces of the nanodiamonds due to photo-ionization of substitutional nitrogen (which forms a deep electron donor with an ionization energy of approximately 1.7~eV) and thus nitrogen atoms in the focal volume of the 532~nm excitation laser may be photo-ionized and release electrons\cite{Tamarat06}. Other studies of blinking NV emission in HPHT nanodiamonds attribute this effect to a charge transfer via electron tunneling between the diamond and the glass/borosilicate substrate.\cite{Bradac13} In both bulk and nanodiamonds, blinking in the microsecond time scale, has been recently ascribed to a photo-conversion process of NV itself, between a bright (very likely the negative charge state) and a dark (very likely the neutral charge state) state \cite{Waldherr11, Siyushev12}. However recently blinking occurring in the ms time scale has been shown due to a two-photon conversion process involved in the charge conversion of NV, achieved with very low excitation power\cite{Aslam13}. By contrast, in our experiment blinking was initiated at high excitation power and the rate of blinking was observed to increase with an increase in the excitation power.


Our findings here seem to suggest another possible mechanism that could cause additional non-radiative decay and blinking. Since we observed increased nonradiative emission, it may appear that some new energy levels are arising in the bandgap of the emitter which results in increased nonradiative relaxations of emitter from its excited state to the ground state. These new non-radiative decay channels lead to a decrease of the fluorescence intensity and at the same time to a decrease of the excited state lifetime of the NV center. Any quenching due to the graphite phases around the diamond surface\cite{Smith10} should have been removed after the initial oxidation steps\cite{Gaebel12}. During successive oxidation steps, oxygen complexes (the diamond surface is expected to be surrounded by highly electronegative oxygen complexes resulting from the oxidation process \cite{Osswald06}) become closer to the NV in the nanodiamonds after their size reduction. We speculate that these oxygen complexes could provide acceptor levels on the surface of the particle, which can lead to additional non-radiative decay channels responsible for a reduction of the total lifetime and the fluorescence counts. In addition, a photo-induced fluctuation of these non-radiative decay channels could also be responsible for subsequent blinking or photo-instability. 

In summary our study shows for the first time that for single NV in nanodiamond, the spontaneous emission follows a local density of states-related suppression for larger diamond crystals (from 300 nm to 150 nm). For smaller crystals additional non-radiative decay channels emerge in the spontaneous decay (from around 150 nm to 30 nm). Our experimental results clearly reveal the advent of new non-radiative decay mechanisms which are responsible for the suppression of NV emission in smaller nanodiamonds. This may well be the reason for the observed large suppression and variability in the NV quantum efficiencies in small nanodiamond crystals \cite{Mohtashami13}. Therefore for spontaneous emission control experiments, NVs in small nanodiamonds should be properly selected for their quantum efficiencies before being used as probes for the local density of states (LDOS). Our work also reveals the possibility that more than one type of blinking inducing mechanism may be present in NV centers in diamond, as was recently shown in quantum dots.\cite{Galland11} Our single centers did not survive all the way to sub-10~nm crystals. However, given that the deterioration of NV emission in the nanodiamonds occurs just before the NV is completely removed from the crystal, we expect that similar surface effects could be observed by repeating the same experiment starting from smaller nanodiamonds and reducing the size down to single-digit scale, and that our results are relevant to the previous observations of complex emission dynamics in those very small diamonds.

\vspace{3mm}
\begin{acknowledgments}
This work was supported in part by the ARC Centre of Excellence for Engineered Quantum Systems(EQuS) (project number CE110001013). Substrate grids were fabricated by the ANFF funded Optofab at Macquarie University (B. Johnston). 
\end{acknowledgments}

\vspace{3mm}

\bibliography{size_reduct_NV}

\end{document}